\documentclass[journal]{IEEEtran}
\usepackage{algorithm}% http://ctan.org/pkg/algorithm
\usepackage{algpseudocode}% http://ctan.org/pkg/algorithmicx
\usepackage{amstext,amsmath,amssymb,exscale,mathtools,bm,gensymb,cite}
\usepackage{psfrag}
\usepackage[none]{hyphenat}
\usepackage{tabularx}
\usepackage[shortlabels]{enumitem}
\usepackage{xcolor}
\usepackage{caption,subcaption,siunitx}
\usepackage[nolist,nohyperlinks]{acronym}
\usepackage{scalerel}
\usepackage{makecell}
\usepackage{graphicx}
\usepackage{hyperref}

\usepackage[font=small]{caption}
\begin{acronym}[MPC] % Give the longest label here so that the list is nicely aligned

    \acro{5G}{fifth generation} 
     
    \acro{6G}{sixth generation} 
    
    \acro{AR}{auto-regressive}

    \acro{ADR}{achievable data rate}
    
    \acro{AoA}{angle of arrival}
    
    \acro{AoD}{angle of departure}

    \acro{AWGN}{additive white Gaussian noise}
    
    \acro{B5G}{beyond fifth generation}

    \acro{BS}{base station}

    \acro{BPSK}{binary phase-shift keying}
    
    \acro{IRS}{intelligent reflecting surface}

    \acro{mmWave}{millimeter Wave}
    
    \acro{MIMO}{multiple input multiple output}

    \acro{CSI}{channel state information}

    \acro{DFT}{discrete Fourier transform}

    \acro{HOSVD}{high order single value decomposition}
    
    \acro{LOS}{line of sight}
    
    \acro{MMSE}{minimum mean squared error}

    \acro{NMSE}{normalized mean square error}

    \acro{NLOS}{non-line of sight}
    
    \acro{SNR}{signal-to-noise ratio}

    \acro{SVD}{singular value decomposition}

    \acro{SE}{spectral efficiency}
    
    \acro{UE}{user equipment}
    
    \acro{ULA}{uniform linear array}

    \acro{URA}{uniform rectangular array}

    \acro{ALS}{alternating least squares}
    
    \acro{KF}{Kronecker factorization}
    
    \acro{KRF}{Khatri-Rao factorization}

    \acro{LS}{least squares}    
    
\end{acronym}

\usepackage{tikz,tikz-3dplot}
\usetikzlibrary{angles, calc, positioning, quotes, intersections,fit,decorations.pathreplacing}
\tdplotsetmaincoords{70}{120}

\renewcommand{\baselinestretch}{0.95}
\usepackage{lipsum}  

\pdfoutput=1

\begin{document}

\begin{acronym}[MPC] % Give the longest label here so that the list is nicely aligned
	
	\acro{5G}{fifth generation} 
	
	\acro{6G}{sixth generation} 
	
	\acro{AR}{auto-regressive}
	
	\acro{ADR}{achievable data rate}
	
	\acro{AoA}{angle of arrival}
	
	\acro{AoD}{angle of departure}
	
	\acro{AWGN}{additive white Gaussian noise}
	
	\acro{B5G}{beyond fifth generation}
	
	\acro{BS}{base station}
	
	\acro{BPSK}{binary phase-shift keying}
	
	\acro{IRS}{intelligent reflecting surface}
	
	\acro{mmWave}{millimeter Wave}
	
	\acro{MIMO}{multiple input multiple output}
	
	\acro{CSI}{channel state information}
	
	\acro{DFT}{discrete Fourier transform}
	
	\acro{HOSVD}{high order single value decomposition}
	
	\acro{LOS}{line of sight}
	
	\acro{MMSE}{minimum mean squared error}
	
	\acro{NMSE}{normalized mean square error}
	
	\acro{NLOS}{non-line of sight}
	
	\acro{SNR}{signal-to-noise ratio}
	
	\acro{SVD}{singular value decomposition}
	
	\acro{SE}{spectral efficiency}
	
	\acro{UE}{user equipment}
	
	\acro{ULA}{uniform linear array}
	
	\acro{URA}{uniform rectangular array}
	
	% Algorithms
	
	\acro{ALS}{alternating least squares}
	
	\acro{KF}{Kronecker factorization}
	
	\acro{KRF}{Khatri-Rao factorization}
	
	\acro{LS}{least squares}    
	
\end{acronym}

\renewcommand{\baselinestretch}{0.90}

\title{PARAFAC-Based Time-Varying Channel Estimation for IRS-Aided Communications}

\author{Kenneth B. A. Benício, André L. F. de Almeida, Bruno Sokal, Fazal-E-Asim, Behrooz Makki, and Gabor Fodor}

%\date{}

\maketitle
 
\begin{abstract}
    This paper proposes a tensor-based parametric channel estimation technique for \acs{IRS}-assisted communication systems with time-varying channel parameters. We exploit the multidimensional structure of the received signal by developing a $3$rd-order PARAFAC tensor model that is solved by employing the iteratively \acs{ALS} algorithm. Our simulation results show that the proposed approach provides enhanced performance in terms of \acs{NMSE} of the concatenated channel compared to the competing solutions by capitalizing on the intrinsic tensor structure of the received signal without increasing the computational complexity of the channel estimation.
\end{abstract}


\section{Introduction}  
    With the increasing number of users due to the implementation of the \ac{5G} and \ac{B5G}, there is a growing demand for network resources in terms of either energy or spectrum usage \cite{7839836}. Most of the recent works focus on solving problems of cooperative relaying, beamforming design, or resource allocation at either the \ac{UE} or the \ac{BS} \cite{9427093,9355403}. However, the propagation environment remains an unknown factor not accounted for in formulating these problems \cite{wu2021intelligent}. \\
    \indent \textcolor{black}{In this context, \acf{IRS} has been explored as one of the possible technologies for integration into \ac{B5G} and \ac{6G} wireless networks aimed at realizing crucial functionalities such as system integration and data transmission \cite{9464918}. This interest arises from the potential of \ac{IRS} to enhance network coverage and establish an intelligent propagation environment \cite{zheng2022survey, pan2022overview, astrom2024ris}. An \ac{IRS} is a two-dimensional panel composed of many passive reflecting elements and a smart controller, which controls the reflecting elements to independently change the phase shifts of impinging electromagnetic waves to smartly maximize the \ac{SNR} by adding them constructively or destructively at the intended receiver \cite{wu2021intelligent}. As \ac{IRS} operates as a passive structure, its power consumption primarily arises from the operation of the smart controller, which lacks the capability to process or amplify incoming signals.} \\
    \indent \textcolor{black}{Since the \ac{IRS} usually is composed of passive elements, it is unable to employ signal processing techniques to the impinging signals, meaning that channel estimation techniques must be carried out at the end nodes of the network, either the \ac{BS} or the \ac{UE}, by transmitting pilot signals according to a training protocol. Several works have already addressed this problem, as the ones mentioned in \cite{de2021channel,AraujoSAM2020,benicio2023channel,benicio2023tensorbased}.} \\
    \indent In \cite{de2021channel},  to perform supervised parameter estimation, a tensor approach is employed, in which the decoupling of the \acs{BS}-\ac{IRS} and \ac{IRS}-\acs{UE} channels is achieved by modeling the received signal at the \ac{BS} as a $3$rd order PARAFAC tensor. In our previous work \cite{benicio2023channel}, we proposed a $3$th order Tucker to solve the problem of parameter estimation of (quasi)-static channels in the context of an \ac{IRS}-aided \ac{MIMO} system. The problem was solved using the \ac{HOSVD} and \acf{ALS} algorithms, with no increase in computational complexity compared to competing methods. Also, in \cite{benicio2023tensorbased}, we proposed a scenario for data-aided tracking where all involved channels were time-varying. This was solved by employing tensor methods on a two-stage parameter estimation framework. In this work, we consider that the \ac{IRS} and the \ac{BS} are in fixed positions, which is more realistic, meaning that the \ac{BS}-\ac{IRS} channel is static. \\
    \indent In this paper, we propose a tensor modeling for the received signal at the \ac{BS} that models the signal as a $3$rd order PARAFAC tensor to solve the parameter estimation problem by employing the \ac{ALS} algorithm \cite{comon2009tensor} in the context of a single time-varying channel. Unlike our approach in \cite{benicio2023tensorbased}, this new solution solves the problem by developing a model that exploits the static nature of the channel between the \ac{BS} and the \ac{IRS}. Usually, this is an acceptable consideration since the \ac{BS} and the \ac{IRS} are deployed in fixed positions. We also provide system design recommendations for our proposed solution and discuss its computational complexity of the proposed solution. Our simulation results show that the proposed tensor-based solution achieves better performance in terms of \acf{NMSE} of the concatenated channel as the classical \ac{LS} filter and the proposed \ac{KRF} in \cite{de2021channel} without increasing the computational complexity. \\
    \indent \textit{Notation}: Scalars, vectors, matrices, and tensors are represented as $a, \boldsymbol{a}, \boldsymbol{A}$, and $\mathcal{A}$. Also, $\boldsymbol{A}^{*}$, $\boldsymbol{A}^{\text{T}}$, $\boldsymbol{A}^{\text{H}}$, and $\boldsymbol{A}^{\dagger}$ stand for the conjugate, transpose, Hermitian, and pseudo-inverse of a matrix $\boldsymbol{A}$, respectively. The $j$th column of $\boldsymbol{A} \in \mathbb{C}^{I \times J}$ is denoted by $\boldsymbol{a}_{j} \in \mathbb{C}^{I \times 1}$. The operator vec$(\cdot)$ transforms a matrix into a vector by stacking its columns, e.g., $\text{vec}(\boldsymbol{A}) = \boldsymbol{a} \in \mathbb{C}^{IJ \times 1}$, while the unvec$(\cdot)_{{I \times J}}$ operator undo the operation. The operator D$(\cdot)$ converts a vector into a diagonal matrix,  $\text{D}_j(\boldsymbol{B})$ forms a diagonal matrix $R \times R$ out of the $j$th row of $\boldsymbol{B} \in \mathbb{C}^{J \times R}$. Also, $\boldsymbol{I}_{N}$ denotes an identity matrix of size $N \times N$. The symbols $\otimes$ and $\diamond$ indicate the Kronecker and Khatri-Rao products.

\section{System Model}
    \begin{figure}[t!]
        \centering
        \begin{tikzpicture}[scale=0.825, every node/.style={scale=0.825}]
            \begin{scope}[
                box1/.style={draw=black, thick, rectangle,rounded corners, minimum height=0.5cm, minimum width=0.5cm}]
                % The first IRS
                \node (IRS) at (-3.25,0.75) {$\text{IRS}$};
                \draw[black,dashed,fill=red!30] (-4.75,-2.5) rectangle (-1.75,.5);
                \node[box1, fill=green!30] (c1) at (-4.25,0) {};
                \node[box1, fill=green!30, right=.125cm of c1] (c2) {};
                \node[box1, fill=green!30, right=.125cm of c2] (c3) {};
                \node[box1, fill=green!30, right=.125cm of c3] (c4) {};
                \node[box1, fill=green!30, below=.125cm of c4] (c5) {};
                \node[box1, fill=green!30, left=.125cm of c5] (c6) {};
                \node[box1, fill=green!30, left=.125cm of c6] (c7) {};
                \node[box1, fill=green!30, left=.125cm of c7] (c8) {};
                \node[box1, fill=green!30, below=.125cm of c8] (c9) {};
                \node[box1, fill=green!30, right=.125cm of c9] (c10) {};
                \node[box1, fill=green!30, right=.125cm of c10] (c11) {};
                \node[box1, fill=green!30, right=.125cm of c11] (c12) {};
                \node[box1, fill=green!30, below=.125cm of c12] (c13) {};
                \node[box1, fill=green!30, left=.125cm of c13] (c14) {};
                \node[box1, fill=green!30, left=.125cm of c14] (c15) {};
                \node[box1, fill=green!30, left=.125cm of c15] (c15) {};
                
                %% The dual function BS
                \draw[black,fill=blue!30] (-7,-6) -- (-6.5,-6) node[above]{$\text{BS}$} -- (-6,-6) -- (-6.5,-4.5) -- cycle;
                % The antennas
                \draw[line width=0.25mm,black] (-7,-4.5) -- (-6,-4.5) node[midway,above]{$\cdots$};
                \draw[line width=0.25mm,black] (-7,-4.5) -- (-7,-4.35);
                \draw[line width=0.25mm,black] (-6,-4.5) -- (-6,-4.35);
                \draw[line width=0.25mm,black] (-7.20,-4.25) -- (-7,-4.35) -- (-6.80,-4.25) -- cycle;
                \draw[line width=0.25mm,black] (-6.20,-4.25) -- (-6,-4.35) -- (-5.80,-4.25) -- cycle;
                
                % Draw UE
                \draw[black,fill=blue!30] (1.25 -2.05,-7.5+1.5) rectangle (2.25-2.05,-6.5+1.5) node[pos=.5] {$\text{UE}$};
                % The antennas
                \draw[line width=0.25mm,black] (1.25-2.05,-6.5+1.5) -- (2.25-2.05,-6.5+1.5) node[midway,above]{$\cdots$};
                \draw[line width=0.25mm,black] (1.25-2.05,-6.5+1.5) -- (1.25-2.05,-6.35+1.5);
                \draw[line width=0.25mm,black] (2.25-2.05,-6.5+1.5) -- (2.25-2.05,-6.35+1.5);
                \draw[line width=0.25mm,black] (1.05-2.05,-6.25+1.5) -- (1.25-2.05,-6.35+1.5) -- (1.45-2.05,-6.25+1.5) -- cycle;
                \draw[line width=0.25mm,black] (2.05-2.05,-6.25+1.5) -- (2.25-2.05,-6.35+1.5) -- (2.45-2.05,-6.25+1.5) -- cycle;
    
                % Drawing the arrows
                %BS to IRS
                \draw[line width=0.5mm,black,<-] (-6.5,-3.85) -- (-5,-1) node[midway, left, rotate=+0]{$\boldsymbol{G}$};
                %IRS to UE
                \draw[line width=0.5mm,black,<-] (-1.5,-1) -- (-0.25,-3.85) node[midway, right, rotate=+0]{$\boldsymbol{H}_{k}$};
                %Direct link
                \draw[line width=0.5mm,red!50,dashed,<->] (-5.75,-5.5) -- (-1,-5.5) node[midway, below, rotate=+0]{$\text{Unavailable}$};
            \end{scope}
        \end{tikzpicture}
        \caption{Proposed \acs{IRS}-assisted \acs{MIMO} system scenario.}
        \label{fig:system_model}
    \end{figure}
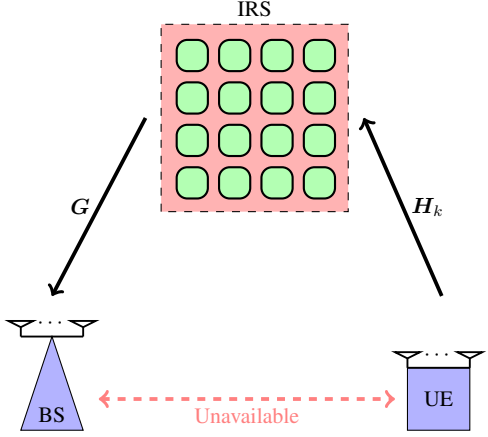
    \begin{figure}[t!]
        \centering
        \begin{tikzpicture}[scale=0.825, every node/.style={scale=0.65}]
            \begin{scope}[
                box1/.style={draw=black, thick, rectangle, minimum height=0.5cm, minimum width=0.5cm}]
                % The first block
                \draw[black,dashed,red,fill=white] (-4.65,-.85) rectangle (-2,.5);
                \node[box1, fill=green!30] (c1) at (-4.25,0) {$1$};
                \node[box1, fill=green!30, right=.125cm of c1] (c2) {$2$};
                \node[right=.125cm of c2] (c3) {$\cdots$};
                \node[box1, fill=green!30, right=.125cm of c3] (c4) {$T$};
                % The second block
                \draw[black,dashed,red,fill=white] (-1.50,-.85) rectangle (1.15,.5);
                \node[box1, fill=green!30] (c1) at (-1.10,0) {$1$};
                \node[box1, fill=green!30, right=.125cm of c1] (c2) {$2$};
                \node[right=.125cm of c2] (c3) {$\cdots$};
                \node[box1, fill=green!30, right=.125cm of c3] (c4) {$T$};
                % The dots
                \node[right=.35 of c4,circle,draw=black,fill=black,minimum size=1pt,scale=0.55] (T1) {};
                \node[right=.15 of T1,circle,draw=black,fill=black,minimum size=1pt,scale=0.55] (T2) {};
                \node[right=.15 of T2,circle,draw=black,fill=black,minimum size=1pt,scale=0.55] (T3) {};
                % The third block
                \draw[black,dashed,red,fill=white] (2.35,-.85) rectangle (5,.5);
                \node[box1, fill=green!30] (c1) at (2.75,0) {$1$};
                \node[box1, fill=green!30, right=.125cm of c1] (c2) {$2$};
                \node[right=.125cm of c2] (c3) {$\cdots$};
                \node[box1, fill=green!30, right=.125cm of c3] (c4) {$T$};
                % The arrows
                \draw[line width=0.15mm,black,<->] (-4.65,-1) -- (-2.,-1) node[midway, below, rotate=+0]{time-varying IRS-UE channel $\boldsymbol{H}_{k}$};
                \draw[line width=0.15mm,black,<->] (-4.65,0.75) -- (5.,0.75) node[midway, above, rotate=+0]{quasi-static BS-IRS channel $\boldsymbol{G}$};
                \draw[line width=0.15mm,black,<->] (-4.45,-0.5) -- (-2.25,-0.5) node[midway, below, rotate=+0]{$1$};
                \draw[line width=0.15mm,black,<->] (-1.3,-0.5) -- (1,-0.5) node[midway, below, rotate=+0]{$2$};
                \draw[line width=0.15mm,black,<->] (2.5,-0.5) -- (4.85,-0.5) node[midway, below, rotate=+0]{$K$};
            \end{scope}
        \end{tikzpicture}
        \caption{Time-domain transmission protocol.}
        \label{fig:transmission_protocol}
    \end{figure}
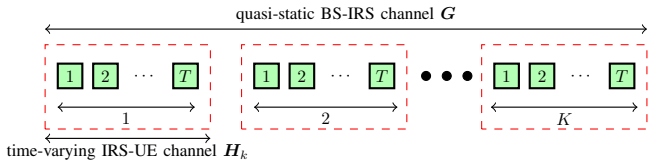
    We consider an uplink \ac{IRS}-assisted \ac{MIMO} scenario with a \ac{BS} equipped with $M$ receiver antennas, which receives a signal from a \ac{UE} equipped with $Q$ transmit antennas \textit{via} a passive \ac{IRS} with $N$ reflecting elements as shown in Fig. \ref{fig:system_model}. The transmission protocol contains $K+1$ blocks each of length of $T$ symbol periods, as shown in Fig~\ref{fig:transmission_protocol}. The received signal is given by 
    \begin{align}
        \boldsymbol{y}_{k,t} &= \boldsymbol{G} \text{D}(\boldsymbol{s}_{t}) \boldsymbol{H}_{k} \boldsymbol{z}_{t} + \boldsymbol{v}_{k,t} \in \mathbb{C}^{M \times 1}, \label{eq:signal_model_1}
    \end{align}
    where $\text{D}(\boldsymbol{s}_{t})$ is the \ac{IRS} phase-shift matrix, $\boldsymbol{z}_{t}$ is the pilot sequence and  $\boldsymbol{v}_{k,t}$ is the \ac{AWGN} vector with $t \in \{1, \cdots, T\}$. We assume that the \ac{IRS}-\ac{UE} link changes faster due to mobility while the \ac{BS}-\ac{IRS} remains static. Specifically, the \ac{IRS}-\ac{UE} channel $\boldsymbol{H}_{k}$ changes between blocks, while the \ac{BS}-\ac{IRS} channel $\boldsymbol{G}$ remains constant during $K+1$ blocks. Assuming a \ac{mmWave} scenario,  we adopt a multipath channel model \cite{heath2016overview} for the involved channels. We can express these channel matrices as follows 
    \begin{align}
        \boldsymbol{G} &= \sum^{L_{1}}_{l_{1} = 1} \alpha^{(l_{1})} \boldsymbol{a}_{\text{r}_{\text{x}}}(\mu^{(l_{1})}_{\text{bs}})  \boldsymbol{b}^{\text{(irs)}\text{H}}_{\text{t}_{\text{x}}}(\mu^{(l_{1})}_{\text{irs}_{{\text{D}}}},\psi^{(l_{1})}_{\text{irs}_{\text{D}}}) , \\
        \boldsymbol{H}_{k} &= \sum^{L_{2}}_{l_{2} = 1} \beta^{(l_{2})}_{k} \boldsymbol{b}^{\text{(irs)}}_{\text{r}_{\text{x}}}(\mu^{(l_{2})}_{\text{irs}_{{\text{A}}}},\psi^{(l_{2})}_{\text{irs}_{\text{A}}}) \boldsymbol{a}_{\text{t}_{\text{x}}}^{\text{H}}(\mu^{(l_{2})}_{\text{ue}}),
    \end{align}
    where $L_{1}$ and $L_{2}$ are the number of directions for channels $\boldsymbol{G}$ and $\boldsymbol{H}_{k}$, respectively. The $l$th \ac{BS} steering vector $\boldsymbol{a}_{\text{r}_{\text{x}}}(\mu^{(l_{1})}_{\text{bs}})$ is associated with the spatial frequency $\mu^{(l_{1})}_{\text{bs}} = \pi \text{cos}(\phi^{(l_{1})}_{\text{bs}})$, with $\phi^{(l_{1})}_{\text{bs}}$ being the \ac{AoA}, which can be further written as 
    \begin{align}
        \boldsymbol{a}_{\text{r}_{\text{x}}}(\mu^{(l_{1})}_{\text{bs}}) = \left[1, \cdots, e^{-j (M - 1) \mu^{(l_{1})}_{\text{bs}}}\right]^{\text{T}} \in \mathbb{C}^{M \times 1}.
    \end{align}
    
    Similarly, the $p$th one-dimensional steering vector for the \ac{UE} is $\boldsymbol{a}_{\text{t}_{\text{x}}}(\mu^{(l_{2})}_{\text{ue}})$ having spatial frequency, which is defined as $\mu^{(l_{2})}_{\text{ue}} = \pi \text{cos}(\phi^{(l_{2})}_{\text{ue}})$, 
    with $\phi^{(l_{2})}_{\text{ue}}$ being the \ac{AoD}, and can be written in terms of spatial frequency as 
    \begin{align}
        \boldsymbol{a}_{\text{t}_{\text{x}}}(\mu^{(l_{2})}_{\text{ue}}) = \left[1, \cdots, e^{-j (Q - 1) \mu^{(l_{2})}_{\text{ue}}}\right]^{\text{T}} \in \mathbb{C}^{Q \times 1}.
    \end{align}
    
    At the \ac{IRS}, $\boldsymbol{b}^{(\text{irs})}_{\text{r}_{\text{x}}}(\mu^{(l_{2})}_{\text{irs}_{{\text{A}}}},\psi^{(l_{2})}_{\text{irs}_{\text{A}}})$ is the 2D steering 
    vector with spatial frequencies $\mu^{(l_{2})}_{\text{irs}_{{\text{A}}}} = \pi \text{cos}(\phi^{(l_{2})}_{\text{irs}_{\text{A}}}) \text{sin} (\theta^{(l_{2})}_{\text{irs}_{\text{A}}})$ and $\psi^{(l_{2})}_{\text{irs}_{\text{A}}} = \pi \text{cos}(\phi^{(l_{2})}_{\text{irs}_{\text{A}}})$, where $\phi^{(l_{2})}_{\text{irs}_{\text{A}}}$ and $\theta^{(l_{2})}_{\text{irs}_{\text{A}}}$ 
    are the azimuth \ac{AoA} and the elevation \ac{AoA}, respectively. This can be further written as the Kronecker product between two steering vectors as
    \begin{align}
        \boldsymbol{b}^{\text{(irs)}}_{\text{r}_{\text{x}}}(\mu^{(l_{2})}_{\text{irs}_{{\text{A}}}},\psi^{(l_{2})}_{\text{irs}_{\text{A}}}) = \boldsymbol{b}^{\text{(irs)}}_{\text{r}_{\text{x}}}(\mu^{(l_{2})}_{\text{irs}_{{\text{A}}}}) \otimes \boldsymbol{b}^{\text{(irs)}}_{\text{r}_{\text{x}}}(\psi^{(l_{2})}_{\text{irs}_{\text{A}}}) \in \mathbb{C}^{N \times 1}.
    \end{align}
    
    The \ac{IRS} transmission steering vector, $\boldsymbol{b}^{\text{(irs)}}_{\text{t}_{\text{x}}}(\mu^{(l_{1})}_{\text{irs}_{{\text{D}}}},\psi^{(l_{1})}_{\text{irs}_{\text{D}}})$, with spatial frequencies defined as $\mu^{(l_{1})}_{\text{irs}_{{\text{D}}}} = \pi \text{cos}(\phi^{(l_{1})}_{\text{irs}_{\text{D}}}) \text{sin} (\theta^{(l_{1})}_{\text{irs}_{\text{D}}})$ and $\psi^{(l_{1})}_{\text{irs}_{\text{D}}} = \pi \text{cos}(\phi^{(l_{1})}_{\text{irs}_{\text{D}}})$, where $\phi^{(l_{1})}_{\text{irs}_{\text{D}}}$ and $\theta^{(l_{1})}_{\text{irs}_{\text{D}}}$ are respectively the azimuth \ac{AoD} and the elevation \ac{AoD}, is given by 
    \begin{align}
        \boldsymbol{b}^{\text{(irs)}}_{\text{t}_{\text{x}}}(\mu^{(l_{1})}_{\text{irs}_{{\text{D}}}},\psi^{(l_{1})}_{\text{irs}_{\text{D}}}) = \boldsymbol{b}^{\text{(irs)}}_{\text{t}_{\text{x}}}(\mu^{(l_{1})}_{\text{irs}_{{\text{D}}}}) \otimes \boldsymbol{b}^{\text{(irs)}}_{\text{t}_{\text{x}}}(\psi^{(l_{1})}_{\text{irs}_{\text{D}}}) \in \mathbb{C}^{N \times 1}.
    \end{align}
    
    The \ac{IRS} phase-shift vector is defined as $\boldsymbol{s}_{t} = \left[e^{j \theta_{1,t}}, \cdots, e^{j \theta_{N,t}}\right]^{\text{T}}  \in \mathbb{C}^{N \times 1}$, where $\theta_{n,t}$ is the phase-shift of the $n$th \ac{IRS} element at the $t$th time slot. Moreover, $\boldsymbol{\alpha} = [\alpha^{(1)}, \cdots, \alpha^{(L_{1})} ]^{\text{T}} \in \mathbb{C}^{L_{1} \times 1}$ and $\boldsymbol{\beta}_{k} = [\beta^{(1)}_{k}, \cdots, \beta^{(L_{2})}_{k}]^{\text{T}} \in \mathbb{C}^{L_{2} \times 1}$ collect the path loss and fading components of the \ac{BS}-\ac{IRS} and \ac{IRS}-\ac{UE} links, respectively. The aging effects are modeled by assuming that $\boldsymbol{\beta}_{k} \in \mathbb{C}^{L_{2} \times 1}$ vary according to a first-order \ac{AR} process defined as \cite{fodor2021performance}
    \begin{align}
        \boldsymbol{\beta}_{k}& = \lambda \boldsymbol{\beta}_{k} + \boldsymbol{\xi}_{k}, k = \{1, \cdots, K\}, \label{eq:beta}
    \end{align}
    \noindent where $\boldsymbol{\xi}_{k} \sim \mathcal{CN}(\boldsymbol{0},(1 -\lambda^{2}) \boldsymbol{I}_{L_{2}}) \in \mathbb{C}^{L_{2} \times 1}$ is the \ac{AR} process noise term for the \ac{IRS}-\ac{UE} link with $\lambda$ being its correlation coefficient \cite{makki2013feedback}. We can compact the notation for $\boldsymbol{G}$ and $\boldsymbol{H}_{k}$ as follows
    \begin{align}
        \boldsymbol{G} &= \boldsymbol{A}_{\text{r}_{\text{x}}} \boldsymbol{D} (\boldsymbol{\alpha}) \boldsymbol{B}^{\text{H}}_{\text{t}_{\text{x}}} \in \mathbb{C}^{M \times N}, \label{eq:channel_G} \\
        \boldsymbol{H}_{k} &= \boldsymbol{B}_{\text{r}_{\text{x}}} \boldsymbol{D} (\boldsymbol{\beta}_{k}) \boldsymbol{A}_{\text{t}_{\text{x}}}^{\text{H}} \in \mathbb{C}^{N \times Q} \label{eq:channel_H},
    \end{align}
    where $\boldsymbol{A}_{\text{r}_{\text{x}}}$, $\boldsymbol{A}_{\text{t}_{\text{x}}}$, $\boldsymbol{B}_{\text{r}_{\text{x}}}$, and $\boldsymbol{B}_{\text{t}_{\text{x}}}$ are the steering matrices defined as 
    \begin{align*}
        \boldsymbol{A}_{\text{r}_{\text{x}}} &= \left[\boldsymbol{a}_{\text{r}_{\text{x}}}(\mu^{(1)}_{\text{bs}}), \cdots, \boldsymbol{a}_{\text{r}_{\text{x}}}(\mu^{(L_{1})}_{\text{bs}}) \right] \in \mathbb{C}^{M \times L_{1}}, \\
        \boldsymbol{A}_{\text{t}_{\text{x}}} &= \left[\boldsymbol{a}_{\text{t}_{\text{x}}}(\mu^{(1)}_{\text{ue}}), \cdots, \boldsymbol{a}_{\text{t}_{\text{x}}}(\mu^{(L_{2})}_{\text{ue}}) \right] \in \mathbb{C}^{Q \times L_{2}}, \\
        \boldsymbol{B}_{\text{r}_{\text{x}}} &= \left[\boldsymbol{b}^{(\text{irs})}_{\text{r}_{\text{x}}}(\mu^{(1)}_{\text{irs}_{{\text{A}}}},\psi^{(1)}_{\text{irs}_{\text{A}}}), \cdots, \boldsymbol{b}^{\text{(irs)}}_{\text{r}_{\text{x}}}(\mu^{(L_{2})}_{\text{irs}_{{\text{A}}}},\psi^{(L_{2})}_{\text{irs}_{\text{A}}}) \right] \in \mathbb{C}^{N \times L_{2}}, \\
        \boldsymbol{B}_{\text{t}_{\text{x}}} &= \left[\boldsymbol{b}^{(\text{irs})}_{\text{t}_{\text{x}}}(\mu^{(1)}_{\text{irs}_{{\text{D}}}},\psi^{(1)}_{\text{irs}_{\text{D}}}), \cdots, \boldsymbol{b}^{\text{(irs)}}_{\text{t}_{\text{x}}}(\mu^{(L_{1})}_{\text{irs}_{{\text{D}}}},\psi^{(L_{1})}_{\text{irs}_{\text{D}}})\right] \in \mathbb{C}^{N \times L_{1}}.
    \end{align*} 
\section{PARAFAC-Based Pilot Signal Design}
    This section describes the proposed tensor-based method for channel parameter estimation exploiting a $3$rd order PARAFAC tensor. Using properties $\text{vec}(\boldsymbol{A} \boldsymbol{B} \boldsymbol{C}) = (\boldsymbol{C}^{\text{T}} \otimes \boldsymbol{A}) \text{vec}(\boldsymbol{B})$ and $\text{vec}(\boldsymbol{A} \text{D} \left(\boldsymbol{b}\right) \boldsymbol{C}) = (\boldsymbol{C}^{\text{T}} \diamond \boldsymbol{A})\boldsymbol{b}$ in the signal model of (\ref{eq:signal_model_1}), yields
        \begin{align}
            \notag \boldsymbol{y}_{k,t} &= \text{vec }(\boldsymbol{I}_{M} \boldsymbol{G} \text{D}(\boldsymbol{s}_{t}) \boldsymbol{H}_{k} \boldsymbol{z}_{t}) + \boldsymbol{v}_{k,t} \in \mathbb{C}^{M \times 1}, \\
            \notag &= (\boldsymbol{z}^{\text{T}}_{t} \otimes \boldsymbol{I}_{M}) \text{vec} (\boldsymbol{G} \text{D}(\boldsymbol{s}_{t}) \boldsymbol{H}_{k}) + \boldsymbol{v}_{k,t}, \\
             &= (\boldsymbol{s}^{\text{T}}_{t} \otimes \boldsymbol{z}^{\text{T}}_{t} \otimes \boldsymbol{I}_{M}) \text{vec}(\boldsymbol{H}^{\text{T}}_{k} \diamond \boldsymbol{G}) + \boldsymbol{v}_{k,t}.
        \end{align}
        and applying again property $\text{vec}(\boldsymbol{A} \boldsymbol{B} \boldsymbol{C}) = (\boldsymbol{C}^{\text{T}} \otimes \boldsymbol{A}) \text{vec}(\boldsymbol{B})$ give us
        \begin{align}
            \notag \boldsymbol{y}_{k,t} &= \text{vec}[(\boldsymbol{z}^{\text{T}}_{t} \otimes \boldsymbol{I}_{M}) (\boldsymbol{H}^{\text{T}}_{k} \diamond\boldsymbol{G}) \boldsymbol{s}_{t}] + \boldsymbol{v}_{k,t}, \\
            &= (\boldsymbol{s}^{\text{T}}_{t} \otimes \boldsymbol{z}^{\text{T}}_{t} \otimes \boldsymbol{I}_{M}) \text{vec}(\boldsymbol{H}^{\text{T}}_{k} \diamond \boldsymbol{G}) + \boldsymbol{v}_{k,t}.
        \end{align}
        
        Collecting the signals during the $T$ symbol periods yields
        \begin{align}
             \boldsymbol{y}_{k} &= \left[ \boldsymbol{y}_{k,1}^{\text{T}}, \cdots, \boldsymbol{y}_{k,T}^{\text{T}} \right]^{\text{T}}, \\
             &= [(\boldsymbol{S} \diamond \boldsymbol{Z})^{\text{T}} \otimes \boldsymbol{I}_{M}] \text{vec}(\boldsymbol{H}^{\text{T}}_{k} \diamond \boldsymbol{G}) + \boldsymbol{v}_{k} \in \mathbb{C}^{M T \times 1}, \\
             &= \boldsymbol{\Omega}\boldsymbol{u}_{k} + \boldsymbol{v}_{k} \in \mathbb{C}^{M T \times 1}, \label{eq:concatenation}
         \end{align} 
         where $\boldsymbol{S} = \left[ \boldsymbol{s}_{1}, \cdots, \boldsymbol{s}_{T} \right] \in \mathbb{C}^{N \times T}$, $\boldsymbol{Z} = \left[ \boldsymbol{z}_{1}, \cdots, \boldsymbol{z}_{T} \right] \in \mathbb{C}^{Q \times T}$ are matrices collecting the \ac{IRS} phase-shifts and pilots, $\boldsymbol{\Omega} = (\boldsymbol{S} \diamond \boldsymbol{Z})^{\text{T}} \otimes \boldsymbol{I}_{M} \in \mathbb{C}^{M T \times M Q N}$,        
        $\boldsymbol{u}_{k} = \textrm{vec}(\boldsymbol{H}^{\text{T}}_{k} \diamond \boldsymbol{G}) \in \mathbb{C}^{M Q N \times 1}$, and $\boldsymbol{v}_{k} = \left[\boldsymbol{v}^{\text{T}}_{k,1}, \cdots, \boldsymbol{v}^{\text{T}}_{k,T} \right]^{\text{T}} \in \mathbb{C}^{M T \times 1}$ is the \ac{AWGN} term.        
        From (\ref{eq:concatenation}), we formulate the following \ac{LS} problem
        \begin{align}
            \boldsymbol{\hat{u}}_{k} = \underset{\boldsymbol{u}_{k}}{\text{arg min}} \left|\left| \boldsymbol{y}_{k} - \boldsymbol{\Omega} \boldsymbol{u}_{k} \right|\right|^2_{2}, \label{eq:least_squares}
        \end{align}
        where the solution requires $T \geq Q N$ and is given by
        \begin{align}
            \boldsymbol{\hat{u}}_{k} &= \boldsymbol{\Omega}^{\dagger} \boldsymbol{y}_{k} \in \mathbb{C}^{MQN \times 1}.
            \label{eq:least_squares_solution}
        \end{align}
        
        Let us define $\boldsymbol{R}_{k} = \textrm{unvec}_{MQ \times N}(\boldsymbol{\hat{u}}_{k}) \approx \boldsymbol{H}^{\text{T}}_{k} \diamond \boldsymbol{G} \in \mathbb{C}^{MQ \times N}$. Using (\ref{eq:channel_G}) and (\ref{eq:channel_H}), while applying property $(\boldsymbol{A} \otimes \boldsymbol{B})(\boldsymbol{C} \diamond \boldsymbol{D}) = (\boldsymbol{AC}) \diamond (\boldsymbol{BD})$, we have
        \begin{align}
            \boldsymbol{R}_{k} &\approx [\boldsymbol{A}_{\text{t}_{\text{x}}}^{*} \text{D} (\boldsymbol{\beta}_{k} ) \boldsymbol{B}^{\text{T}}_{\text{r}_{\text{x}}}] \diamond [\boldsymbol{A}_{\text{r}_{\text{x}}}\text{D} (\boldsymbol{\alpha}) \boldsymbol{B}^{\text{H}}_{\text{t}_{\text{x}}}], \\
            &\approx (\boldsymbol{A}_{\text{t}_{\text{x}}}^{*} \otimes \boldsymbol{A}_{\text{r}_{\text{x}}}) [(\text{D} (\boldsymbol{\beta}_{k}) \boldsymbol{B}^{\text{T}}_{\text{r}_{\text{x}}}) \diamond (\text{D} (\boldsymbol{\alpha}) \boldsymbol{B}^{\text{H}}_{\text{t}_{\text{x}}})], \\
            &\approx  ( \boldsymbol{A}_{\text{t}_{\text{x}}}^{*}  \otimes  \boldsymbol{A}_{\text{r}_{\text{x}}} )  [\text{D} (\boldsymbol{\beta}_{k}) \otimes  \text{D} (\boldsymbol{\alpha})] (\boldsymbol{B}^{\text{T}}_{\text{r}_{\text{x}}}  \diamond  \boldsymbol{B}^{\text{H}}_{\text{t}_{\text{x}}}). \label{eq:combined_channel_1}
        \end{align}
        Let us consider $\boldsymbol{f}_{k} = \boldsymbol{\beta}_{k} \otimes \boldsymbol{\alpha} \in \mathbb{C}^{L_{1} L_{2} \times 1}$, then (\ref{eq:combined_channel_1}) can be rewritten as
        \begin{align}
            \boldsymbol{R}_{k} &\approx \boldsymbol{A} \text{D}(\boldsymbol{f}_{k}) \boldsymbol{B}^{\text{T}} \in \mathbb{C}^{M Q \times N}, \label{eq:combined_channel_2}
        \end{align} 
        where $\boldsymbol{A} = (\boldsymbol{A}_{\text{t}_{\text{x}}}^{*} \otimes \boldsymbol{A}_{\text{r}_{\text{x}}}) \in \mathbb{C}^{M Q \times L_{1} L_{2}}$ and $\boldsymbol{B} = (\boldsymbol{B}^{\text{T}}_{\text{r}_{\text{x}}} \diamond \boldsymbol{B}^{\text{H}}_{\text{t}_{\text{x}}}) \in \mathbb{C}^{N \times L_{1} L_{2}}$ represents the channel geometry information from the \ac{BS}-\ac{UE} and \ac{IRS}, respectively. The collection of matrices $\{\boldsymbol{R}_{1}, \ldots, \boldsymbol{R}_{K}\}$, in (\ref{eq:combined_channel_2}) over all blocks $k \in \{1, \ldots, K\}$ can be arranged as a third-order PARAFAC tensor $\mathcal{R} \in \mathbb{C}^{M Q \times N \times K}$, which can be expanded in terms of a tensor notation as
        \begin{align}
            \mathcal{R} &\approx \mathcal{I}_{3,L_{1} L_{2}} \times_{1} \boldsymbol{A} \times_{2} \boldsymbol{B} \times_{3} \boldsymbol{F}^{\text{T}} \in \mathbb{C}^{M Q \times N \times K}, \label{eq:combined_channel_tensor}
        \end{align}
        where $\boldsymbol{F} = [\boldsymbol{f}_{1}, \cdots, \boldsymbol{f}_{K}] \in \mathbb{C}^{L_{1} L_{2} \times K}$ is the combined pathloss across all $K$ blocks. The matrix unfoldings of $\mathcal{R}$ are given by
        \begin{align}
            \left[\mathcal{R}\right]_{(1)} &= \boldsymbol{A} \left(\boldsymbol{F}^{\text{T}} \diamond \boldsymbol{B}\right)^{\text{T}} \in \mathbb{C}^{M Q \times N K}, \\
            \left[\mathcal{R}\right]_{(2)} &= \boldsymbol{B} \left(\boldsymbol{F}^{\text{T}} \diamond \boldsymbol{A}\right)^{\text{T}} \in \mathbb{C}^{N \times M Q K}, \\
            \left[\mathcal{R}\right]_{(3)} &= \boldsymbol{F}^{\text{T}} \left(\boldsymbol{B} \diamond \boldsymbol{A}\right)^{\text{T}} \in \mathbb{C}^{K \times M Q N}.
        \end{align}
        
        Consequently the estimation of $\boldsymbol{A}$, $\boldsymbol{B}$, and $\boldsymbol{F}$ consists of solving the following optimization problem
        \begin{align}
            \left\{\hat{\boldsymbol{A}}, \hat{\boldsymbol{B}}, \hat{\boldsymbol{F}}\right\} = \underset{\boldsymbol{A}, \boldsymbol{B}, \boldsymbol{F}} {\text{arg min}} \left|\left|  \mathcal{R} -  \mathcal{I}_{3,L_{1} L_{2}} \times_{1} \boldsymbol{A} \times_{2} \boldsymbol{B} \times_{3} \boldsymbol{F}^{\text{T}} \right|\right|^{2}_{\text{F}}, \label{eq:optimization_problem_tensor}
        \end{align}    
        which can be performed by means of the \ac{ALS} algorithm (summarized in Algorithm \ref{alg:ALS}) \cite{comon2009tensor,de2016overview}.
        
        \subsection{\ac{ALS} Channel Parameter Estimation}
            \begin{algorithm}[!t]
                \caption{Alternating least squares \label{alg:ALS}}
                \begin{algorithmic}[1]
                    \Require{Tensor $\mathcal{R}$, maximum number of iterations $i_{\text{max}}$, convergence threshold $\delta$.}
                    \State{Initialize randomly $\boldsymbol{A}$, $\boldsymbol{B}$, and $\boldsymbol{F}$ at iteration $i = 0$.}
                    \While{$||e(i) - e(i-1)|| \geq \delta$ and $i < i_{\text{max}}$}
                        \State{Find a least squares estimate of $\boldsymbol{A}$ as}
                        \begin{align*}
                            \hat{\boldsymbol{A}} &=  \left[\mathcal{R}\right]_{(1)} \left[(\hat{\boldsymbol{F}}^{\text{T}} \diamond \hat{\boldsymbol{B}})^{\text{T}}\right]^{\dagger}.
                        \end{align*}
                        \State{Find a least squares estimate of $\boldsymbol{B}$ as}
                        \begin{align*}
                            \hat{\boldsymbol{B}}  &=  \left[\mathcal{R}\right]_{(2)} \left[(\hat{\boldsymbol{F}}^{\text{T}} \diamond \hat{\boldsymbol{A}})^{\text{T}}\right]^{\dagger}.
                        \end{align*}
                        \State{Find a least squares estimate of $\boldsymbol{F}$ as}
                        \begin{align*}
                            \hat{\boldsymbol{F}} &=  \left(\left[\mathcal{R}\right]_{(3)} \left[(\hat{\boldsymbol{B}} \diamond \hat{\boldsymbol{A}})^{\text{T}}\right]^{\dagger}\right)^{\text{T}}.
                        \end{align*}
                        \State{Repeat \textit{until} convergence.}
                    \EndWhile
                    \State{\textbf{return} $\hat{\mathcal{R}} \approx \mathcal{I}_{3,L_{1} L_{2}} \times_{1} \hat{\boldsymbol{A}} \times_{2} \hat{\boldsymbol{B}} \times_{3} \hat{\boldsymbol{F}}^{\text{T}}$.}
                \end{algorithmic}
            \end{algorithm}
            In this scenario, the algorithm consists of estimating $\boldsymbol{A}$, $\boldsymbol{B}$, and $\boldsymbol{F}$ in an alternating way by iteratively solving the following cost functions
            \begin{align}
                \hat{\boldsymbol{A}} &= \underset{\boldsymbol{A}}{\text{arg min}} \left|\left|  \left[\mathcal{R}\right]_{(1)} - \boldsymbol{A} (\boldsymbol{F}^{\text{T}} \diamond \boldsymbol{B})^{\text{T}}  \right|\right|^{2}_{\text{F}}, \label{eq:least_squares_mode_1} \\
                \hat{\boldsymbol{B}} &= \underset{\boldsymbol{B}} {\text{arg min}} \left|\left|  \left[\mathcal{R}\right]_{(2)} - \boldsymbol{B} (\boldsymbol{F}^{\text{T}} \diamond \boldsymbol{A})^{\text{T}} \right|\right|^{2}_{\text{F}}, \label{eq:least_squares_mode_2} \\
                \hat{\boldsymbol{F}} &= \underset{\boldsymbol{F}^{\text{T}}}{\text{arg min}} \left|\left|  \left[\mathcal{R}\right]_{(3)} - \boldsymbol{F}^{\text{T}} (\boldsymbol{B} \diamond \boldsymbol{A})^{\text{T}}  \right|\right|^{2}_{\text{F}}, \label{eq:least_squares_mode_3}
            \end{align} 
            with the solutions for (\ref{eq:least_squares_mode_1})-(\ref{eq:least_squares_mode_3}) being respectively given by
            \begin{align}
                \hat{\boldsymbol{A}} &=  \left[\mathcal{R}\right]_{(1)} \left[(\boldsymbol{F}^{\text{T}} \diamond \boldsymbol{B})^{\text{T}}\right]^{\dagger} \in \mathbb{C}^{M Q \times L_{1} L_{2}}, \label{eq:mode_1_estimation} \\
                \hat{\boldsymbol{B}}  &=  \left[\mathcal{R}\right]_{(2)} \left[(\boldsymbol{F}^{\text{T}} \diamond \boldsymbol{A})^{\text{T}}\right]^{\dagger} \in \mathbb{C}^{N \times L_{1} L_{2}}, \label{eq:mode_2_estimation} \\
                \hat{\boldsymbol{F}} &=  \left(\left[\mathcal{R}\right]_{(3)} \left[(\boldsymbol{B} \diamond \boldsymbol{A})^{\text{T}}\right]^{\dagger}\right)^{\text{T}} \in \mathbb{C}^{L_{1} L_{2} \times K}, \label{eq:mode_3_estimation}
            \end{align}
            with each solution requiring that 
            \begin{align}
                L_{1} L_{2} \leq N K, \,\,\, L_{1} L_{2} \leq M Q K, \,\,\, L_{1} L_{2} \leq M Q N.
            \end{align}     
            
            These conditions are necessary to guarantee the existence of the \ac{LS} estimates of $\boldsymbol{A}$, $\boldsymbol{B}$, and $\boldsymbol{F}$, respectively, by ensuring that the pseudoinverses on Equations (\ref{eq:mode_1_estimation})-(\ref{eq:mode_3_estimation}) are well defined. The proposed \ac{ALS} to solve the problem in (\ref{eq:optimization_problem_tensor}) consists of three iterative and alternating update steps that follow the \ac{LS} solutions in (\ref{eq:mode_1_estimation})-(\ref{eq:mode_3_estimation}). The reconstruction error is minimized to a one-factor matrix at each update by fixing the remaining matrices to their previous estimation. This procedure is repeated until convergence, which happens when the reconstruction error of consecutive iterations, given by $e(i) = ||\mathcal{R} - \hat{\mathcal{R}}(i)||^{2}_{\text{F}}$, achieves $||e(i) - e(i-1)|| \leq \epsilon$ and $\epsilon$ is the threshold parameter with $\hat{\mathcal{R}}(i)$ being the estimated tensor fit model at the $i$th iteration. We initialize the factor matrices randomly, and the convergence threshold is set to $\epsilon = 10^{-5}$.  
        \subsection{Computational complexity}
             In Table \ref{tab:computational_complexity}, we describe the computational complexity for the selected benchmark algorithms, the \ac{LS} at (\ref{eq:least_squares_solution}), the \ac{KRF} from \cite{de2021channel}, and our proposed PARAFAC \ac{ALS} algorithm. Consider that the pseudo-inverse of a matrix $\mathbf{A} \in \mathbb{C}^{I \times J}$, with $I > J$, and its rank-$R$ \ac{SVD} approximation have complexities $\mathcal{O}(I J^{2})$ and $\mathcal{O}(I J R)$ \cite{kishore2017literature}, respectively. The \ac{KRF} \cite{de2021channel} estimates the combined channel $\boldsymbol{R}_{k} = \boldsymbol{H}^{\text{T}}_{k} \diamond \boldsymbol{G}$ by finding estimates of both $\boldsymbol{G}$ and $\boldsymbol{H}_{k}$ that solves a set of $N$ rank-one approximations using the \ac{SVD} along $K$ blocks. Regarding the proposed algorithm, the \ac{ALS} computes $3$ pseudo-inverses (\ref{eq:mode_1_estimation})-(\ref{eq:mode_3_estimation}) along $\text{\acs{ALS}}_{\text{iter}}$ iterations until convergence over $K$ blocks.
            \begin{table}[!t]
                \centering
                \caption{Computational complexity: \acs{LS}, \acs{KRF}, and \acs{ALS}.} \label{tab:computational_complexity}
                \begin{tabular}{|c|c|}
                    \hline
                    Algorithm                                       & Computational Complexity              \\ \hline
                    \acs{LS} (\ref{eq:least_squares_solution})    & $\mathcal{O}(K (M Q N)^{3})$            \\ \hline
                    \acs{KRF} \cite{de2021channel}                  & $\mathcal{O}(K M Q N)$                  \\ \hline
                    \acs{ALS} (Alg. \ref{alg:ALS})                               & $\mathcal{O}(K M Q N \text{ALS}_{\text{iter}} (L_{1} L_{2})^{2} (1 + \frac{K}{N} + \frac{K}{MQ} ))$                                                                                     \\ \hline
                \end{tabular}
            \end{table}
        
\section{Simulation Results}
\begin{figure}[!t]
        \centering
        \includegraphics[width=.8\linewidth]{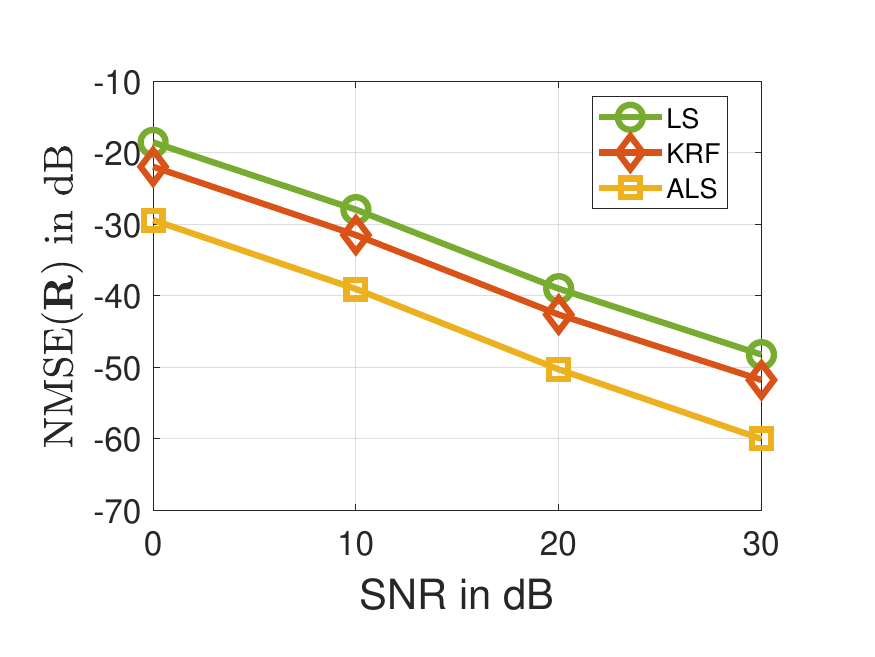} 
        \caption{Performance evaluation in terms of \ac{NMSE} for the competing algorithms, the \ac{LS} filter in (\ref{eq:least_squares_solution}) and the \ac{KRF} \cite{de2021channel}, and for the proposed \ac{ALS} solution in Algorithm \ref{alg:ALS}.}
        \label{fig:nmse}
    \end{figure}

\begin{figure}[!t]
        \centering
        \includegraphics[width=.8\linewidth]{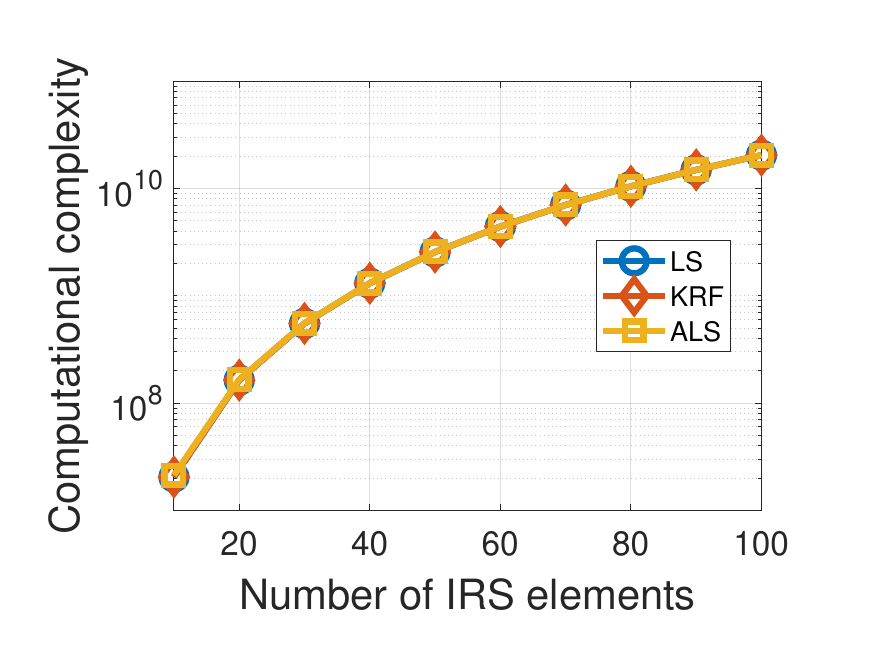}
        \caption{Computational complexity cost of the competing solutions and the proposed \ac{ALS} solution in Alg \ref{alg:ALS} according to the Table \ref{tab:computational_complexity}.}
        \label{fig:computational_complexity_case_1}
    \end{figure}

\begin{figure}[!t]
        \centering
        \begin{subfigure}{.475\textwidth}
            \centering
            \includegraphics[width=.8\linewidth]{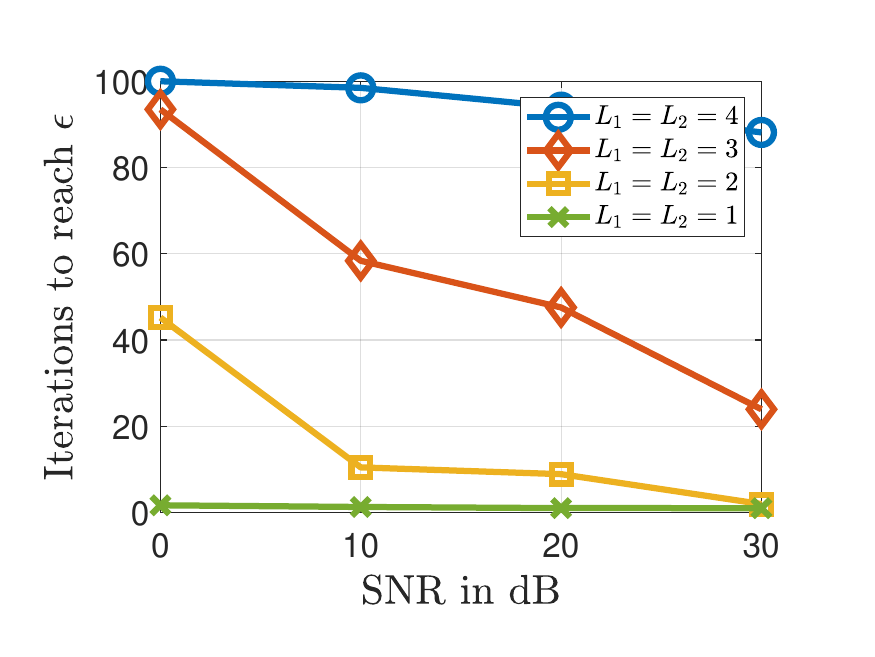}
            \caption{Convergence as function of the \ac{SNR}  for different numbers of paths  $L_{1}$ and $L_{2}$.}
            \label{fig:iter_case_1}
        \end{subfigure}
        \begin{subfigure}{.475\textwidth}
            \centering
            \includegraphics[width=.8\linewidth]{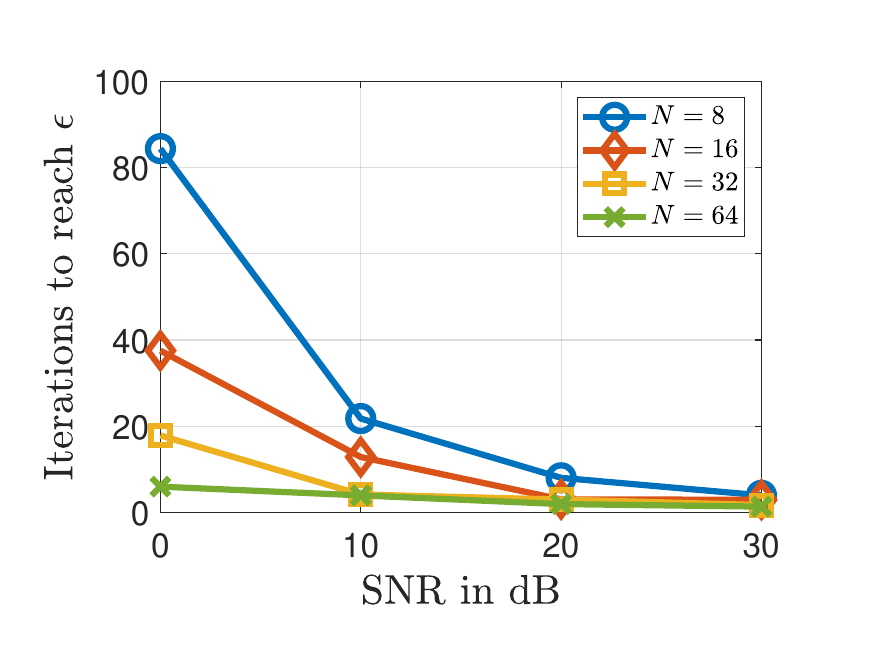}
            \caption{Convergence as function of the \ac{SNR} as a function of $N$.}
            \label{fig:iter_case_2}
        \end{subfigure}
        \caption{Performance evaluation in terms of the required number of iterations for the proposed \ac{ALS} solution in (\ref{alg:ALS}).}
    \end{figure}

    We evaluate the performance of the proposed tensor-based algorithm by comparing it again with the reference parameter estimation method based on the \ac{KRF} \cite{de2021channel}. The pilot signal matrix $\boldsymbol{Z} \in \mathbb{C}^{Q \times T}$ is designed as a Hadamard matrix, while a \ac{DFT} is adopted for the \ac{IRS} phase-shift matrix $\boldsymbol{S}$. The angular parameters $\phi^{(l_{1})}_{\text{bs}}$ and $\phi^{(l_{2})}_{\text{ue}}$ are randomly generated from a uniform distribution between $[-\pi, \pi]$ while the \ac{IRS} elevation and azimuth angles of arrival and departure are randomly generated from a uniform distribution between $[-\pi/2, \pi/2]$. The fading coefficients $\boldsymbol{\alpha}$ and $\boldsymbol{\beta}_{k}$ are modeled as independent Gaussian random variables $\mathcal{CN}(0,1)$. The parameter estimation accuracy is evaluated in terms of the \ac{NMSE} given as 
    \begin{align}
        \text{\ac{NMSE}}(\boldsymbol{R}) = \mathbb{E} \left\{\frac{\left|\left|\boldsymbol{R}^{(e)}_{k} - \hat{\boldsymbol{R}}^{(e)}_{k}\right|\right|^{2}_{\text{F}}}{\left|\left|\boldsymbol{R}^{(e)}_{k}\right|\right|^{2}_{\text{F}}} \right\},
    \end{align}
    where $\boldsymbol{\hat{R}}_{k}$ is the estimated tensor fit reconstructed with (\ref{eq:combined_channel_tensor}) at the $e$th run, with $E = 10^4$ being the number of Monte Carlo trials. Unless otherwise stated, the system parameters are $\{M = 4, Q = 4, L_{1} = 2, L_{2} = 2, N = 16, T = 64, K = 5,\text{ and } \delta = 0.75\}$.  \\
    \indent In Fig. \ref{fig:nmse}, we evaluate the \ac{NMSE} performance associated with the estimation of the combined channel $\boldsymbol{R}_{k} = \boldsymbol{H}^{\text{T}}_{k} \diamond \boldsymbol{G}$, as a function of the training \ac{SNR} to compare selected competing algorithms,  the classical \ac{LS} filter as in  (\ref{eq:least_squares_solution}), the state-of-the-art \ac{KRF} \cite{de2021channel}, and our proposed solution in Algorithm \ref{alg:ALS}. We observe that the proposed solution in Algorithm \ref{alg:ALS} outperforms the \ac{LS} filter and the state-of-the-art \ac{KRF} \cite{de2021channel} algorithms by approximately $10$ dB and $7$ dB, respectively. Also, this gain is almost independent of the \ac{SNR}. \textcolor{black}{In the case of the \ac{LS}, the estimation is worse because this solution does not exploit the intrinsic Khatri-Rao structure of the channel, while the \ac{KRF} exploits the separability of the channel structure to refine the estimation process a step further than the \ac{LS}}. In contrast, the proposed \ac{ALS} is an iterative solution that refines the estimations until convergence is declared. Furthermore, we also exploit the geometric structure of the scenario in the proposed solution. Across all considered methods, we observe that the \ac{NMSE} decreases linearly with \ac{SNR} in the log domain, as expected.
     \\ 
    \indent In Fig. \ref{fig:computational_complexity_case_1}, we analyze the computational complexity of the competing algorithms and the proposed solution in Algorithm \ref{alg:ALS}. In this scenario, we fixed all the system parameters except for the number of reflecting elements at the \ac{IRS}, depicted by $N$, according to Table \ref{tab:computational_complexity}. To compute the cost of the \ac{KRF} \cite{de2021channel}, and the \ac{ALS} in Algorithm \ref{alg:ALS}, we take into account the additional cost from the computation of the combined channel parameters by the \ac{LS} filter in (\ref{eq:least_squares_solution}) which is the expensive step of the involved solutions (see Table \ref{tab:computational_complexity}). Moreover, simulations show that the proposed \ac{ALS} algorithm takes approximately $10$ iterations to converge. We observe that the competing \ac{LS} and \ac{KRF} \cite{de2021channel} algorithms have approximately the same cost as the proposed \ac{ALS} solution. 
     \\
    \indent In Fig. \ref{fig:iter_case_1}, we evaluate the number of iterations required by the proposed \ac{ALS} solution in Algorithm \ref{alg:ALS} to accomplish convergence as a function of the \ac{SNR} and the number of channel directions, depicted as $L_{1}$ and $L_{2}$, respectively. We set the target convergence criterion to $\epsilon = 10^{-5}$, which means that convergence is declared when the fit error of the estimated tensor model between consecutive iterations is less than $\epsilon$. As expected, in the low \ac{SNR} region ($< 10$ dB), the \ac{ALS} solution in Algorithm \ref{alg:ALS} takes more iterations to declare convergence as the noise variance is higher. Also, when the total number of components, i.e., the product $L_{1} L_{2}$, grows more iterations are needed to solve the problem in (\ref{eq:optimization_problem_tensor}). On the other hand, at a high SNR regime ($> 20$ dB), the total number of iterations required for achieving the convergence is considerably lower (around $80\%$ reduction to cases $L_{1} L_{2} = 2$ and $L_{1} L_{2} = 3$) however,  when $L_{1} L_{2} = 16$, the proposed \ac{ALS} algorithm does not achieve the convergence criterion under $100$ iterations, i.e., a lower threshold is required or a higher number of iterations at the cost of a higher computational complexity. At the high \ac{SNR} region ($> 20$ dB), the required number of iterations is considerably lower than in most scenarios. However, in the scenario where $L_{1} L_{2} = 16$ components, the proposed algorithm still fails to converge within $100$ iterations. \\
    \indent In Fig. \ref{fig:iter_case_2}, we take the scenario of Fig. \ref{fig:iter_case_1} for the case where we have $L_{1} L_{2} = 4$ components and evaluate the impact of the number of reflecting elements of the \ac{IRS} in the convergence of the \ac{ALS} solution at Algorithm \ref{alg:ALS}. We observe that, as the number of reflecting elements $N$ increases, fewer iterations are needed for convergence, which is linked to the \ac{LS} filter in (\ref{eq:least_squares_solution}) since, if $N$ increases, we can sense the channel longer to achieve a more precise acquisition of the combined \ac{CSI}. Furthermore, we observe that in this scenario, all cases converge in the high \ac{SNR} region ($> 20$ dB), regardless of the number of reflecting elements.     
\section{Conclusions}
    This paper proposes a tensor-based channel estimation algorithm for a single time-varying channel in \ac{IRS}-assisted systems. Unlike our work in \cite{benicio2023channel}, we assumed that the channel \ac{BS}-\ac{IRS} remains quasi-static, whereas the channel \ac{IRS}-\ac{UE} has time-varying fading components. In contrast, the parametric structure of the scenario is considered to remain approximately constant. We have proposed a tensor model for the reflected signal from the \ac{IRS} that employs a $ 3$-rd-order PARAFAC tensor structure. To solve the \ac{CSI} acquisition in this scenario, we derive an \ac{ALS} solution according to Alg. \ref{alg:ALS}. The parameter estimation accomplishes the estimation of only the overall channel \ac{UE}-\ac{IRS}-\ac{BS}.
    \bibliographystyle{ieeetr}
    \bibliography{addons/bibliography}
\end{document}